\documentclass[prl,twocolumn,showpacs,superscriptaddress]{revtex4}

\usepackage{graphicx}
\usepackage{dcolumn}
\usepackage{amsmath}

\newcommand{\LaSr}{La$_{1-x}$Sr$_x$MnO$_3$}

\newcommand{\etal}{\textit{et al.}}

\newcommand{\LaSre}{La$_{0.875}$Sr$_{0.125}$MnO$_3$}

\newcommand{\Mnthree}{Mn$^{3+}$}
\newcommand{\Mnfour}{Mn$^{4+}$}

\begin{document}
\title{Observation of a Griffiths Phase in paramagnetic \LaSr}

\author{J.~Deisenhofer}
\affiliation{EP V, Center for Electronic Correlations and Magnetism,
University of Augsburg, 86135 Augsburg, Germany}

\author{D.~Braak}
\affiliation{Theoretical Physics II, Institute for Physics,
University of Augsburg, 86135 Augsburg, Germany}

\author{H.-A.~Krug von Nidda}
\affiliation{EP V, Center for Electronic Correlations and Magnetism,
University of Augsburg, 86135 Augsburg, Germany}

\author{J.~Hemberger}
\affiliation{EP V, Center for Electronic Correlations and Magnetism,
University of Augsburg, 86135 Augsburg, Germany}

\author{R.~M.~Eremina}
\affiliation{E.~K.~Zavoisky Physical-Technical Institute, 420029
Kazan, Russia}

\author{V.~A.~Ivanshin}
\affiliation{Kazan State University, 420008 Kazan, Russia }

\author{A.~M.~Balbashov}
\affiliation{Moscow Power Engineering Institute, 105835 Moscow,
Russia}

\author{G.~Jug}
\affiliation{Dipartimento di Fisica e Matematica, Universita'
dell'Insubria, 22100 Como, Italy}

\author{A.~Loidl}
\affiliation{EP V, Center for Electronic Correlations and Magnetism,
University of Augsburg, 86135 Augsburg, Germany}

\author{T.~Kimura}
\affiliation{Department of Applied Physics, University of Tokyo,
Tokyo 113-8656, Japan}

\author{Y.~Tokura}
\affiliation{Department of Applied Physics, University of Tokyo,
Tokyo 113-8656, Japan}

\date{\today}

\begin{abstract}
We report on the discovery of a novel triangular phase regime in the
 system \LaSr\;by means of electron spin resonance and
magnetic susceptibility measurements. This phase is characterized by
the coexistence of ferromagnetic entities within the globally
paramagnetic phase far above the magnetic ordering temperature. The
nature of this phase can be understood in terms of Griffiths
singularities arising due to the presence of correlated quenched
disorder in the orthorhombic phase.
\end{abstract}



\maketitle

Recently, the considerable influence of \textit{quenched disorder}
on the phase complexity in manganite systems and the appearance of
phenomena like colossal magnetoresistance (CMR) has been unraveled
both experimentally \cite{Akahoshi03,Nakajima04} and theoretically
\cite{Dagotto01,Burgy04}. Within the context of quenched-disorder
scenarios the existence of a Griffiths-like \cite{Griffiths69}
temperature scale $T_G$ above the magnetic ordering temperature
$T_C$ has been predicted and linked to CMR
\cite{Burgy01,Dagotto05,Salamon02}. Moreover, the competition
between charge-ordered antiferromagnetic (AFM) and metallic
ferromagnetic (FM) phases appears to be a significant factor for the
rich phase diagrams of these systems \cite{Dagotto05}, and the
persistence of nanoscale inhomogeneities in the paramagnetic (PM)
regime has been reported early on \cite{Teresa97}.

\begin{figure}[t]
\begin{center}
\includegraphics[width=65mm,clip]{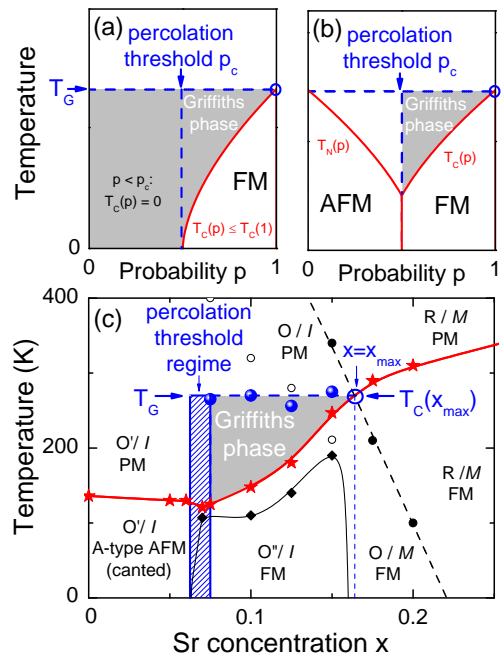}
\end{center}
\caption{(a) $T$-$p$ diagram for the dilute FM Ising model
\cite{Griffiths69}. (b) Conjectured schematic $T$-$p$ diagram of the
GP arising in a $\pm J$ random Ising model due to the competition of
FM/AFM clusters;  (c) observed Griffiths phase boundaries within the
established $T$-$x$ phase diagram of LSMO \cite{Paraskevopoulos00}.
The intersection (open circle) of $T_G$ (spheres) with the magnetic
boundary $T_C$ (stars) coincides with the phase transition from the
orthorhombic ($O$) to rhombohedral ($R$) structure ($I$ = insulator,
$M$= metal). Data for $x$=0.06 and 0.07 were taken from
\cite{Hennion98} and \cite{Dubunin03}, respectively. Lines are drawn
to guide the eye.} \label{phadiaGriff}
\end{figure}

Below $T_G$ the quenched disordered system is in between the
completely disordered PM high-temperature regime and the
magnetically ordered state. This phase regime is usually referred to
as the \textit{Griffiths phase} (GP) \cite{Bray87}, based on
Griffiths' seminal treatment of the effects of quenched randomness
on the magnetization of a dilute Ising ferromagnet
\cite{Griffiths69}. Griffiths showed that essential singularities
would develop in a temperature region $T_C(p)<T<T_G$, where $p$
denotes the disorder parameter, $T_C(p)$ the disorder-dependent FM
ordering temperature (Fig.~\ref{phadiaGriff}a) and $T_G$ a new
temperature scale corresponding to $T_C(1)$, the Curie temperature
of the undiluted system with $p=1$. Further studies showed the
importance of correlated disorder in generating and enhancing the
new singularities \cite{McCoy68,Shan87,Vojta03}, but a systematic
study of the Griffiths phenomenon in a competing 3D two-phase
situation (e.g.\;AFM/FM) is still not available at present. The
impact of Griffiths scenarios for disorder physics is evidenced by
its evocation for such challenging physical problems as the
non-Fermi liquid behavior in Kondo systems \cite{Castro-Neto98} and
the properties of magnetic semiconductors \cite{Galitski04}. To
date, however, an entire GP, i.e.\;a globally PM regime
characterized by the temperature boundaries $T_G$ and $T_C$ and a
well-defined disorder parameter $p$, had not been identified
experimentally.

Here we report the discovery of an entire GP in single crystals of
the paradigm system \LaSr\;(LSMO), demonstrating the impact of
quenched disorder in manganites. Using electron spin resonance (ESR)
and magnetic susceptibility measurements, we clearly identify a
triangular phase regime limited by the Sr concentration $x_c\sim
0.07$, the Griffiths temperature scale $T_G\sim 270$~K, and the FM
transition temperature up to a maximal Sr concentration
$x_{\mathrm{max}}\sim 0.16$ (Fig.~\ref{phadiaGriff} (c)).
Furthermore, we propose that the appearance of Griffiths-phase
regimes can be expected for many other manganite systems and mapped
out by ESR which is a local magnetic probe and particularly
sensitive in the PM regime.

ESR measurements were performed with a Bruker CW spectrometer at 9.4
GHz and 34 GHz. Susceptibilities were measured with a SQUID
magnetometer (Quantum design). Details on the experimental setup and
crystal growth have been published elsewhere
\cite{Urushibara95,Ivanshin00,Deisenhofer03}.

\begin{figure}[h]
\begin{center}
\includegraphics[width=60mm,clip]{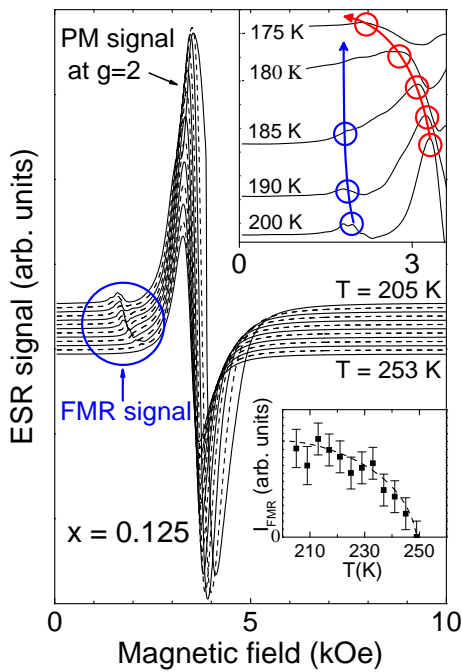}
\end{center}
\caption{ESR spectra for $x=0.125$ for $205 \leq T\leq$ 253~K with
the magnetic field applied within the easy $ac$-plane. Upper inset:
evolution of the spectra towards $T_{\rm C}$. Lower inset:
$T$-dependence of the FMR intensity. Lines are to guide the eye.}
\label{spectra0125}
\end{figure}

In Fig.~\ref{spectra0125} we show ESR spectra in \LaSre \ in the PM
regime above $T_C=180$~K. The spectra do not only consist of a PM
signal due to the majority of Mn$^{3+}$ and Mn$^{4+}$ spins
\cite{Ivanshin00}, but also exhibit an intriguing FM resonance (FMR)
signal at lower resonance fields. A rough estimate obtained by
comparing the FMR and PM signals shows that the fraction of spins
contributing to the FMR is $\lesssim 1\%$. The PM resonance signal
and its anisotropy in the orbitally ordered phase have been analyzed
in detail previously \cite{Deisenhofer03,Deisenhofer02,Kochelaev03}.
Here, the focus is on the FMR signal, which emerges from the signal
of the PM Mn$^{3+}$/Mn$^{4+}$ background ($g=2$) at $T\simeq 260$~K,
far above $T_C=180$~K \cite{Paraskevopoulos00}, and then shifts
towards lower resonance fields indicating an increase of the local
magnetic fields in the sample. This shift corresponds to the
$T$-dependence of the FM magnetization~\cite{Gurevich96}.
Concomitantly, the intensity of the FMR (estimated via its
peak-to-peak linewidth $\Delta H_{pp}$ and its amplitude $A$ as
$\Delta H_{pp}^2\cdot A$) first clearly increases and then saturates
with decreasing temperatures, as shown in the lower inset of
Fig.~\ref{spectra0125}. This behavior excludes a superparamagnetic
origin of the signal, which would result in a $T$-dependence
according to the Langevin function (for temperatures under
consideration, a Curie-Weiss (CW) like increase) instead of a
saturation.

\begin{figure}[h]
\begin{center}
\includegraphics[width=60mm,clip]{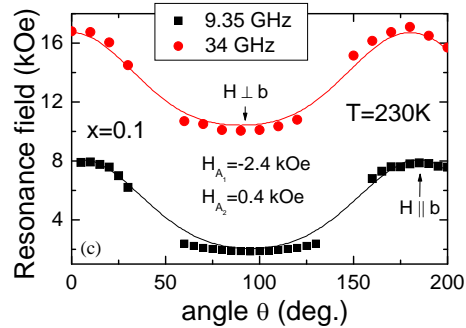}
\end{center}
\caption{Orientation dependence of the resonance field of the FMR at
9.35 GHz and 34 GHz for $x=0.1$ at 230~K. Fitting curves were
obtained by assuming an easy-plane ($\perp b$) anisotropy
\cite{Gurevich96}.} \label{2D0125plot}
\end{figure}

On approaching $T_{\rm C}$ the PM line broadens, shifts to lower
resonance fields, and finally merges with the FMR (upper inset of
Fig.~\ref{spectra0125}). Moreover, the FMR signal exhibits a
pronounced easy-plane anisotropy with respect to the $ac$-plane
($Pnma$) of FM superexchange coupling which is shown in
Fig.~\ref{2D0125plot} for a sample with $x=0.1$ at 230~K. The lack
of data at about 3.3~kOe and 12~kOe is due to the fact that the FMR
signal cannot be resolved anymore when passing through the PM signal
at $g=2$. Fitting the angular dependence of the resonance field
measured at 9.4 GHz and 34 GHz by taking into account first and
second order anisotropy fields $H_{A_1}$ and $H_{A_2}$
\cite{Gurevich96} results in $H_{A_1}=-2.4$~kOe and
$H_{A_2}=0.4$~kOe. However, demagnetization effects due to a
plate-like shape of the FM domains would produce the same kind of
anisotropy and, therefore, a distinction between these two sources
of anisotropy is not possible \cite{Gurevich96}.

These FMR signals were observed in the PM regime above $T_C$ in
single crystals with Sr concentrations $x = 0.075, 0.1, 0.125,
0.15$. They all separate from the PM signal below 270~K indicating a
temperature scale $T_G$ above $T_C$ which is almost independent of
$x$, and they all exhibit the same anisotropy as shown in
Fig.\ref{2D0125plot}. Remarkably, no such additional FMR signals
could be identified above $T_C$ for samples with $x> 0.175$, which
already exhibit a FM metallic ground state.  Moreover, for $x\leq
0.05$ no FMR was observed in the PM regime, suggesting the existence
of a lower Sr concentration threshold $x_c$ with $0.05<x_c<0.075$.
The fact that the phenomenon is not observable outside this
concentration range confirms the intrinsic nature of the FMR
signals.

\begin{figure}[h]
\begin{center}
\includegraphics[width=55mm,clip]{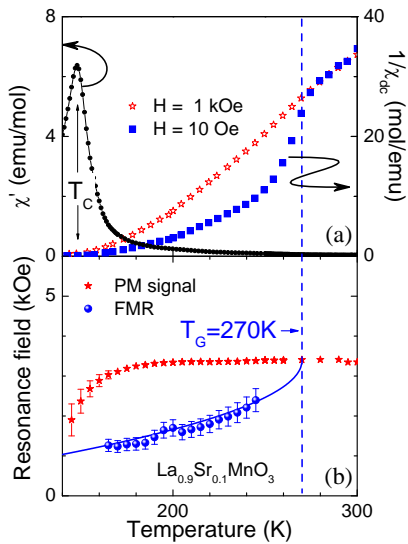}
\end{center}
\caption{$T$-dependence of (a) the ac susceptibility ($H_{ac}=1$~Oe,
$f_{ac}=10$~Hz) and the inverse dc susceptibility at 10 Oe and 1
kOe, and (b) the resonance fields of the FMR and the PM signal for
$x = 0.1$. The solid line describes the FMR shift $\propto(1-
T/T_G)^{1/2}$. The external magnetic field was applied within the
easy $ac$-plane.} \label{chires01}
\end{figure}

In Fig.~\ref{chires01} we show susceptibility data for a sample with
$x=0.1$ (upper panel) together with the $T$-dependence of the
resonance fields of both the FMR and the PM signal (lower panel).
For a large applied magnetic field (1 kOe) the FM component is
hidden in the PM contribution and a CW law is found throughout the
PM regime, while in small magnetic fields (10 Oe) clear deviations
from a CW behavior are observed below $T_G=270$~K
(Fig.~\ref{chires01}a), in agreement with the evolution of the FMR
intensity (lower inset of Fig.~\ref{spectra0125}).

This temperature agrees very well with the extrapolation of the FMR
shift with respect to $g=2$ using a mean-field saturation behavior
$\propto (1-T/T_G)^{1/2}$ (Fig.~\ref{chires01}b). Plotting the
obtained temperatures $T_G(x)$ within the \mbox{$T$-$x$}~phase
diagram of LSMO\;(Fig.~\ref{phadiaGriff}c)
\cite{Paraskevopoulos00,Dabrowski01,Klingeler02}, we obtain a novel
phase regime within the PM region with an almost constant upper
temperature boundary \mbox{$T_G(x)\simeq T_C(x\sim 0.16)$}.

Having identified the new phase regime in low doped LSMO, we will
now discuss its interpretation in the context of quenched disorder
as an enhanced GP becoming observable due to the competition of two
ordering phases \cite{Burgy01}. The source of disorder is the random
substitution of La$^{3+}$ by ions with different size and valence,
such as Sr or Ca. The probability $p(x)$ for the existence of a FM
bond increases with $x$, because the increasing number of
\Mnthree-\Mnfour\;pairs enhances the double-exchange (DE) driven FM
interaction. Due to the static Jahn-Teller (JT) distortion of the
Mn$^{3+}$ ions the non-JT active \Mnfour\;ions and the FM bonds can
be regarded as fixed within the lattice (quenched disorder).


The lower bound $T_G$ for the non-analyticity of the magnetization
is identified with the appearance of the FMR signals at $T\simeq
270$~K. From the intersection of the $T_G$ boundary with the
magnetic ordering boundary, the regular ferromagnet corresponding to
$p=1$ is found to be at $x\sim 0.16$. In Bray's generalization of
Griffiths' concept to FM systems with an arbitrary distribution of
bond strengths, the Griffiths temperature scale $T_G$ is no longer
the critical temperature of the pure FM system with $p=1$ but the
maximal critical temperature among all configurations compatible
with the static nature of disorder \cite{Bray82}. In our case, the
static disorder is annealed above $x_{\mathrm{max}}\sim 0.16$ as a
consequence of the transition from the JT distorted orthorhombic to
the rhombohedral phase, i.e.\;as the random locations of the FM
bonds begin to fluctuate concomitantly with the fluctuating lattice
distortions. Consequently, the Griffiths phenomenon disappears for
$x>0.16$. The importance of this orthorhombic to rhombohedral
transition has recently been reported to be crucial for the
appearance of signatures of correlated clusters above $T_C$
\cite{Kiryukhin04}.

The threshold regime $0.05<x_c<0.075$ derived from the existence of
the FMR in our ESR spectra can be refined by the results of
neutron-diffraction studies in a single crystal with $x=0.07$. At
this concentration the system is mainly in the canted AFM state, but
an estimated 10 \% of the crystal volume is in a FM state
\cite{Dubunin03}. In contrast, for $x=0.06$ no such features were
found \cite{Hennion98}, narrowing the percolation threshold regime
down to $0.06<x_c<0.075$. To estimate $x_c$ for FM bonds via DE we
start with a \textit{sc} lattice of Mn$^{3+}$ ions with lattice
constant $a$. Assuming that upon doping only linear
Mn$^{3+}$-Mn$^{4+}$-Mn$^{3+}$ clusters form a \textit{sc} lattice
(percolation threshold $p_c^{sc}=0.3116$ \cite{Stauffer92}) with
lattice constant $2a$, only $1/8$ of the sites are occupied by
Mn$^{4+}$ ions resulting in a lower bound $x_c^l=p_c^{sc}/8 \simeq
0.04$. Similarly, the staggered-square cluster with two Mn$^{4+}$-
and two Mn$^{3+}$-ions results in a \textit{fcc} lattice
($p_c^{fcc}=0.198$ \cite{Stauffer92}), where half of the sites are
occupied by Mn$^{4+}$-ions. Thus, we derive an upper bound
$x_c^u=p_c^{fcc}/2\simeq 0.1$ and obtain $0.04 < x_c < 0.1$ in good
agreement with experiment.

Keeping in mind that the disorder in LSMO~is quenched within the JT
distorted structure, it becomes clear that the disorder must be of a
correlated nature as assumed in the argument above. In the case of
theoretical models which include correlated disorder, it was found
that Griffiths effects are enhanced \cite{McCoy68,Vojta03}.
Additionally, the existence of AFM bonds in the system and the
resulting two-phase competition have to be considered and, indeed,
Burgy \etal\;\cite{Burgy01,Burgy04} argue on theoretical grounds
that the existence of competing phases stabilizes and enhances FM
Griffiths-like effects. In the presence of AFM clusters, the GP
should be confined to a restricted region of the $T$-$p$~phase
diagram as depicted in Fig.~\ref{phadiaGriff}b.

Correlated disorder was considered by Vojta who studied the
Griffiths phenomenon in a 3D Ising FM with planar defects
\cite{Vojta03}. As expected, the correlated disorder simply makes
the large, rare clusters responsible for the Griffiths singularities
much more likely to occur, thus strengthening considerably the
Griffiths phenomenon. We believe this correlated-disorder
enhancement, together with the two-phase AFM/FM competition, to be
responsible for the observability of the otherwise weak Griffiths
singularities. The critical behavior at the magnetic ordering
temperature in LSMO\;has been studied in detail by Oleaga
\etal\;through thermal diffusivity measurements \cite{Ole04}. These
authors can describe the PM-AFM phase transition for $x<0.1$ by a
3D-Heisenberg-type behavior, whilst for the PM-FM transition for
$x>0.28$ a 3D-Ising model is suggested. In the concentration range
$0.1<x<0.28$ no universal behavior was found in agreement with
theoretical expectations \cite{Vojta03}.

The existence of such a novel phase regime characterized by the FMR
features is not restricted to weakly doped LSMO, but rather
represents a generic feature of manganite systems where the
structural distortions are sufficiently strong to allow for the
bond-disorder to be completely quenched: The coexistence of PM and
FMR signals above $T_C$ has already been reported for some samples
of the layered La$_{2-2x}$Sr$_{1+2x}$Mn$_2$O$_7$ manganites
\cite{Chauvet98,Simon03}, but a complete ESR mapping of this system
has not yet been performed. At the temperature scale where the FMR
signal appears in these compounds, the onset of correlated polaronic
behavior has been reported by neutron-diffraction studies
\cite{Argyriou02,Kiryukhin04}. Recently, Chapman \etal\;reported
similar features for a series of polycrystalline
Ln$_{0.7}$M$_{0.3}$MnO$_3$ samples with Ln=La,Pr,Nd,Sm and
M=Ca,Sr,Ba \cite{Chapman04}, increasing the number of systems where
Griffiths phase regimes can be expected.

The observed value of $T_G=270$~K is naturally explained through the
influence of the structural transition which anneals the disorder
distribution at $x\simeq 0.16$. For $x> 0.16$, the time scale of the
disorder fluctuations is no longer much greater than the dynamical
scale of the order-parameter fluctuations and the Griffiths
singularities disappear. In this regime (rhombohedral phase) a
description based on magnetic polarons coupled to the lattice
distortions is suitable. For $x< 0.16$, the lattice distortions are
essentially frozen providing the fixed background of FM bonds in the
PM insulating state. This quenched nature of the FM bond
configuration agrees with the fact that no significant changes in
the resistivity have been observed at $T_G$ in LSMO
\cite{Mostovshchikova04}. This picture is valid above the transition
to the FM insulator, below which the orbital degrees of freedom come
into play.

In conclusion, we have experimentally identified, to the best of our
knowledge, for the first time an entire Griffiths phase in the
$T$-$x$ phase diagram of LSMO\;by means of ESR and susceptibility
measurements. This phase regime arises as a result of the strong
quenching of the randomly diluted locations of the FM bonds in the
cooperatively JT-distorted structure and can be expected to be a
generic feature in manganites. The enhanced visibility of these
effects in manganites and the characteristic ESR features challenge
further theoretical studies on the base of models which incorporate
both FM/AFM phase competition and quenched disorder.

\begin{acknowledgments}
We thank M.V.~Eremin, K.-H.~H\"{o}ck, T.~Kopp, L.~Svistov, D.~Zakharov
and K.~Ziegler for fruitful discussions. This work was supported by
the BMBF via the contract number VDI/EKM 13N6917 and partly by the
DFG via SFB 484 (Augsburg). Support from MUIR through a COFIN-2003
project (G.J.) and from BRHE/REC007 (R.M.E.) is also acknowledged.
\end{acknowledgments}


\end{document}